\documentclass[final,5p,twocolumn,10pt]{elsarticle}

\usepackage{amssymb}

\usepackage{url}
\usepackage{rotating}

\usepackage{hyperref}
\hypersetup{
 linktocpage=true,
 pdfborderstyle={/S/S/W 1},
 hyperindex=true,
 bookmarks=true,
 bookmarksopen=true,
 bookmarksnumbered=true,
}

\journal{Nuclear Instruments and Methods A}

\journal{NIM}

\begin{document}

\begin{frontmatter}



\title{The camera of the fifth H.E.S.S. telescope.\\Part I: System description.}


\author[lpnhe]{J.~Bolmont\corref{cor}}
\ead{bolmont@in2p3.fr}
\cortext[cor]{Corresponding author. Tel.: +33 144274818, fax. +33~144274638.}
\author[lpnhe]{P.~Corona}
\author[lpnhe]{P.~Gauron\fnref{ipno}}
\fntext[ipno]{Present address: IPNO, Universit\'e Paris Sud, CNRS/IN2P3, 15 rue Georges Cl\'emenceau, F-91406 Orsay Cedex, France}
\author[lpnhe]{P.~Ghislain}
\author[lpnhe]{C.~Goffin}
\author[lpnhe]{L. Guevara Riveros\fnref{ipno}}
\author[lpnhe]{J.-F.~Huppert}
\author[lpnhe]{O.~Martineau-Huynh}
\author[lpnhe]{P.~Nayman}
\author[lpnhe]{J.-M.~Parraud}
\author[lpnhe]{J.-P.~Tavernet}
\author[lpnhe]{F.~Toussenel}
\author[lpnhe]{D.~Vincent}
\author[lpnhe]{P.~Vincent}

\author[apc]{W.~Bertoli}
\author[apc]{P.~Espigat}
\author[apc]{M.~Punch}

\author[irfu]{D.~Besin}
\author[irfu]{E.~Delagnes}
\author[irfu]{J.-F.~Glicenstein}
\author[irfu]{Y.~Moudden}
\author[irfu]{P.~Venault}
\author[irfu]{H.~Zaghia}

\author[lapp]{L.~Brunetti}
\author[lapp]{P.-Y.~David}
\author[lapp]{J-M.~Dubois}
\author[lapp]{A.~Fiasson}
\author[lapp]{N.~Geffroy}
\author[lapp]{I.~Gomes~Monteiro}
\author[lapp]{L.~Journet}
\author[lapp]{F.~Krayzel}
\author[lapp]{G.~Lamanna}
\author[lapp]{T.~Le~Flour}
\author[lapp]{S.~Lees}
\author[lapp]{B.~Lieunard}
\author[lapp]{G.~Maurin}
\author[lapp]{P.~Mugnier} 
\author[lapp]{J-L.~Panazol}
\author[lapp]{J.~Prast}

\author[llr]{L.-M.~Chounet}
\author[llr]{B.~Degrange}
\author[llr]{E.~Edy}
\author[llr]{G.~Fontaine}
\author[llr]{B.~Giebels}
\author[llr]{S.~Hormigos}
\author[llr]{B.~Kh\'elifi}
\author[llr]{P.~Manigot}
\author[llr]{P.~Maritaz}
\author[llr]{M.~de~Naurois}

\author[lupm]{M.~Compin}
\author[lupm]{F.~Feinstein}
\author[lupm]{D.~Fernandez}
\author[lupm]{J.~Mehault}
\author[lupm]{S.~Rivoire}
\author[lupm]{S.~Royer}
\author[lupm]{M.~Sanguillon}
\author[lupm]{G.~Vasileiadis}

\address[lpnhe]{LPNHE, Universit\'e Pierre et Marie Curie Paris 6, Universit\'e Denis Diderot Paris 7, CNRS/IN2P3, 4 Place Jussieu, F-75252 Paris Cedex 5, France}

\address[apc]{APC, AstroParticule et Cosmologie, Universit\'{e} Paris Diderot, CNRS/IN2P3, CEA/Irfu, Observatoire de Paris, Sorbonne Paris Cit\'{e}, 10, rue Alice Domon et L\'{e}onie Duquet, F-75205 Paris Cedex 13, France}

\address[irfu]{CEA Saclay, DSM/IRFU, F-91191 Gif-Sur-Yvette Cedex, France}

\address[lapp]{LAPP, Universit\'{e} de Savoie, CNRS/IN2P3, F-74941 Annecy-le-Vieux, France}

\address[llr]{LLR, Ecole Polytechnique, CNRS/IN2P3, F-91128 Palaiseau, France}

\address[lupm]{LUPM, Universit\'e Montpellier 2, CNRS/IN2P3, CC 72, Place Eug\`ene Bataillon, F-34095 Montpellier Cedex 5, France}

\begin{abstract}
In July 2012, as the four ground-based gamma-ray telescopes of the H.E.S.S. (\textit{High Energy Stereoscopic System}) array reached their tenth year of operation in Khomas Highlands, Namibia, a fifth telescope took its first data as part of the system. This new Cherenkov detector, comprising a $614.5\,\rm{m}^2$ reflector with a highly pixelized camera in its focal plane, improves the sensitivity of the current array by a factor two and extends its energy domain down to a few tens of GeV.

The present part I of the paper gives a detailed description of the fifth H.E.S.S. telescope's camera, presenting the details of both the hardware and the software, emphasizing the main improvements as compared to previous H.E.S.S. camera technology.
\end{abstract}

\begin{keyword}
High Energy Stereoscopic System \sep H.E.S.S. \sep camera \sep electronics \sep hardware \sep software \sep calibration


\end{keyword}

\end{frontmatter}



\section{Introduction}
\label{sec:intro}

In 2005, three years after the first H.E.S.S.\,\footnote{\textit{High Energy Stereoscopic System}.} 12m telescope was commissioned and one year after the fourth was installed, the decision was taken by French and German agencies to build a fifth instrument. This new imaging atmospheric Cherenkov telescope was mainly designed to obtain an energy threshold of $\sim30\,\rm{GeV}$\,\footnote{At trigger level, prior to any selection cut. This value is an estimation obtained from simulations. The exact value will depend on observation conditions.}, increasing the sensitivity of the array towards low energies \cite{hess2punch,hess2holler}. In order to achieve this goal, a new fast, high sensitivity and low dead-time 2048 pixel camera was designed to equip the focal plane of a parabolic tessellated mirror of $614.5\,\rm{m}^2$. With this new telescope, the H.E.S.S. project was entering its second phase.

The goal of the H.E.S.S. telescopes is to record and analyze the short and faint Cherenkov light flash created by very high energy gamma-rays when they interact with the atmosphere and produce an extensive electromagnetic shower \cite{history}. This technique, used by several experiments around the world, has allowed the detection and characterization of more than a hundred Galactic and extragalactic objects so far \cite{tev,tevcat}.

Most of the high energy gamma-ray sources studied by H.E.S.S., as well as the hadronic background, have spectra which can be parameterized with a power-law\,\footnote{However, the exact shape of the source spectra (flux, spectral index, presence of an energy cut-off, \ldots) varies from a source to another.} in the range between tens of GeV and TeV energies. This implies that lowering the energy threshold from hundreds of GeV to tens of GeV requires the data acquisition chain to be able to handle much higher trigger rates. The dead-time has also to be reduced. The camera of the fifth H.E.S.S. telescope (simply called ``CT5'' in the following) is a complete new design allowing to meet these challenging constraints: the dead time was lowered to $15\,\mu\rm{s}$ and the trigger rate\,\footnote{L2A rate, see \S\ref{subsec:trigger}.} can reach $5\,\rm{kHz}$.

Since it was designed several years after the first four H.E.S.S. cameras (in the following, the first four telescopes of H.E.S.S. will be referred to as ``CT1--4'') \cite{hess1cam1, hess1cam2, hess1perf}, the new camera benefits from the latest progress in electronics integration: FPGAs\,\footnote{\textit{Field-Programmable Gate Array}.} and FIFO\,\footnote{\textit{First-In First-Out}.} buffers are used extensively and dedicated ASICs\,\footnote{\textit{Application Specific Integrated Circuits}.} were specifically designed for CT5.

The paper is divided in two parts. This first part gives a complete description of the CT5 camera. The main improvements as compared to the CT1--4 technology will be pointed out when relevant. The second part of the paper, to appear later, will deal with the performance of the camera in its nominal working configuration on-site in Namibia.

The present part is organized as follows. In \S\ref{sec:conespm}, the light guides, photomultiplier tubes and very front-end electronics are described. \S\ref{sec:elec} deals with the electronics, giving details on the front-end and the trigger as well as on the safety, slow control and monitoring (\S\ref{sec:slc}). The mechanics of the camera are discussed in \S\ref{sec:meca}, the calibration instrumentation in \S\ref{sec:calib} and embedded software in \S\ref{sec:soft}.

\section{Photomultipliers and light collection}
\label{sec:conespm}

\begin{figure}[t!] 
   \centering
   \includegraphics[width=3.5in]{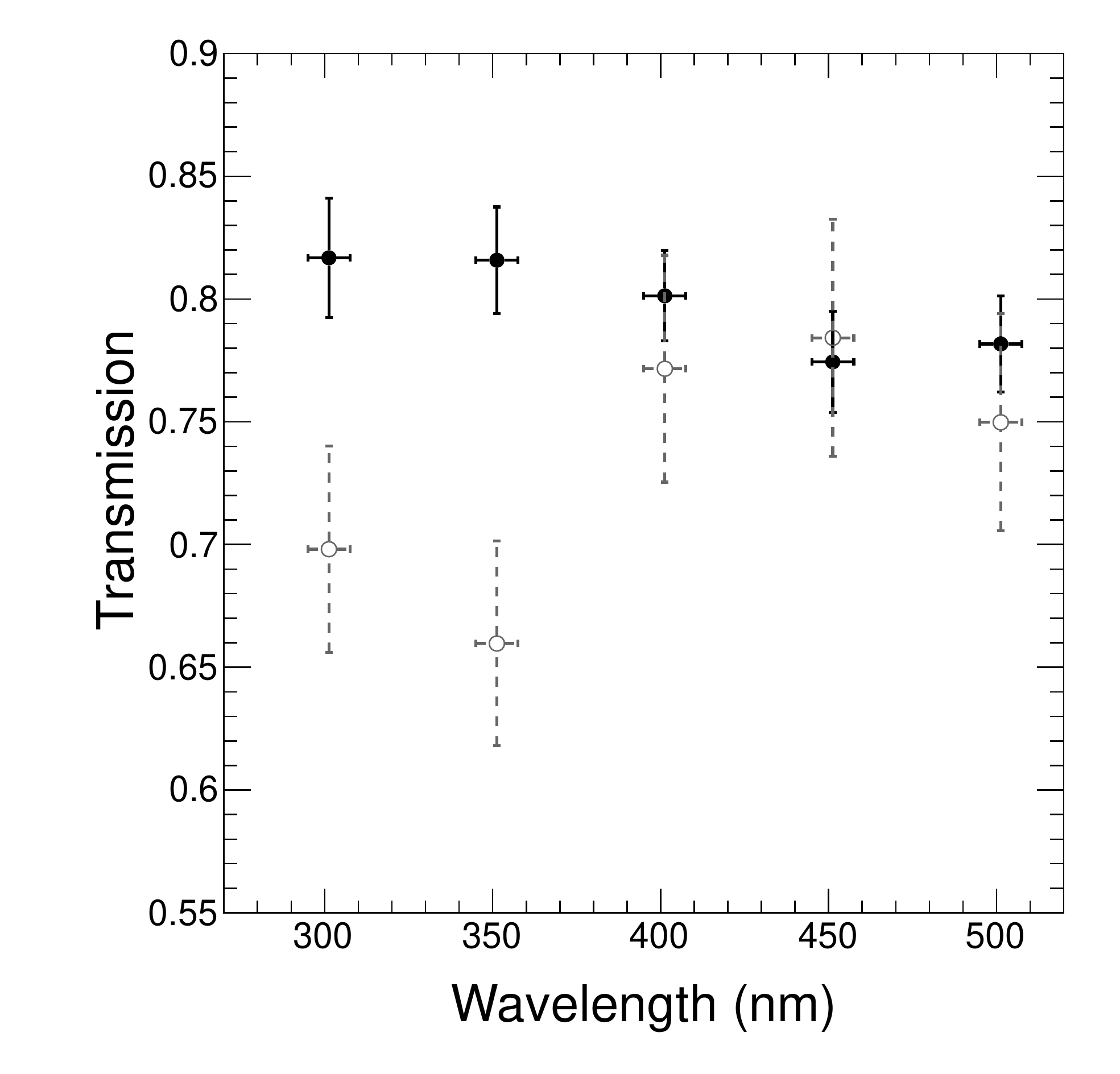} 
\vspace{-0.8cm}
   \caption{Average optical transmission of newly produced CT1--4 (grey open circles) and CT5's (black solid markers) Winston cones as a function of the wavelength for a incoming light direction distribution identical to that of the telescope. Error bars correspond to the r.m.s. over all measurements and do not include systematics (10\%). The difference seen below $400\,\rm{nm}$ comes from the new coating used for CT5 cones.}
   \label{fig:LCtrans}
\end{figure}

The photomultiplier tube (PMT) which was chosen for CT5 is Photonis XP-29600. This PMT is a minor update of the XP-2960 used for CT1--4: the length of the PMT was reduced to decrease the ringing of the anode signal and the average quantum efficiency was improved from 25\% to 30\%. The main characteristics of these PMTs are cited in Table~\ref{tab:h1h2comp}. Each PMT is soldered on a circular base where the resistive divider is located. This base itself is connected to a board where the high voltage (HV) is obtained using a Cockcroft-Walton generator. The high voltage can be set between 0 and $1600\,\rm{V}$ with a $1\,\rm{V}$ resolution. To protect the PMT from high light levels (bright stars, shooting stars), the current (noted HVI in the following) drawn by the PMT is monitored and the high voltage is turned off by the board itself when the current exceeds $200\,\mu\rm{A}$, independently of the slow control and safety management system. The circular base and the HV generation board have been designed and produced by the ISEG Company.

2500 PMTs were tested for use in the CT5 camera, including spares. The tests included gain calibration, measurement of the linearity and of after-pulse rate, as well as photo-cathode uniformity.

Each PMT is equipped with a ``Winston cone'' light guide \cite{Winston} to minimize dead-space between the PMT sensitive areas and to cut down on albedo light (not coming from the mirror). The use of Winston cones in VHE gamma-ray imaging astronomy was introduced in 1994 \cite{Punch-TAMAC}, and is currently used by all imaging Cherenkov cameras. Each Winston cone guides photons which impinge on it with an angle below its ``cut-off'' angle (defined by its geometry) towards the PMT entry window after on average
a single reflection, while reflecting back the photons above this cut-off angle. For a cut-off angle set to be close to the arrival direction of photons coming from the edge of the mirror, this provides almost complete protection from albedo light from the night-sky or from light sources and diffusive reflections from the ground. For CT5, the cut-off angle used for the Winston cones is about $30^\circ$. 

The same physical size of PMT is used as in the first cameras, which combined with the similar $f/D$, allows an identical Winston cone design to be used, for which details can be found in \cite{HESS-opt1}. This incidentally gives a similar Night-Sky Background (NSB) rate per pixel in CT5 as in the first four telescopes (on average $\sim$100\,MHz of NSB photo-electrons), so the PMTs can operate at the same gain. The cones are assembled from injection moulded polycarbonate half-cones with aluminization and MgF$_2$ protection, and have a hexagonal symmetry around their axis, to minimize the dead-space between the cones. The Winston cone entry aperture (at the mirror-facing side) of $42\,\rm{mm}$ flat-to-flat on each hexagon defines the pixel field of view (FoV) of $0.07^\circ$ (compared to $0.16^\circ$ for the first cameras). 

The wall thickness of the cone entry is $0.4\pm0.1\,\rm{mm}$, to minimize the dead area in the focal plane.
In order to reduce their cost and to improve their performance reproducibility, they were produced by an 
industrial process optimised for this production at the company\,\footnote{SAVIMEX, located near Grasse in France.}.

In order to validate the light guide design and to measure each cone after the mass production, the automatic test bench used for the Winston cones of the first cameras was re-used. Measurements of the cut-off angle and of the absolute optical transmission are made with a diffusive source covering the same angular aperture as the CT5 mirror as seen from the camera. 2669 cones have been measured following the production. Fig.~\ref{fig:LCtrans} shows their average absolute optical transmission as a function of the wavelength, defined as the ratio of the amount of light detected by a photo-sensor to that entering the Winston cones. While the transmission is about the same for CT1--4 and CT5 cones above $400\,\rm{nm}$, it has been improved by 15\% below that wavelength in the range detectable by the PMT\@. This difference is due to the new coating used for CT5 Winston cones and should result in an improvement of the global optical efficiency of the telescope by a few percent. As the Winston cone geometry is strictly identical that used in CT1--4, the measured transmission for CT5 as a function of the angle of incidence is identical to that of CT1--4 (see Figure~13 of \cite{HESS-opt1}).

\begin{figure}[t!] 
   \centering
   \includegraphics[width=3.5in]{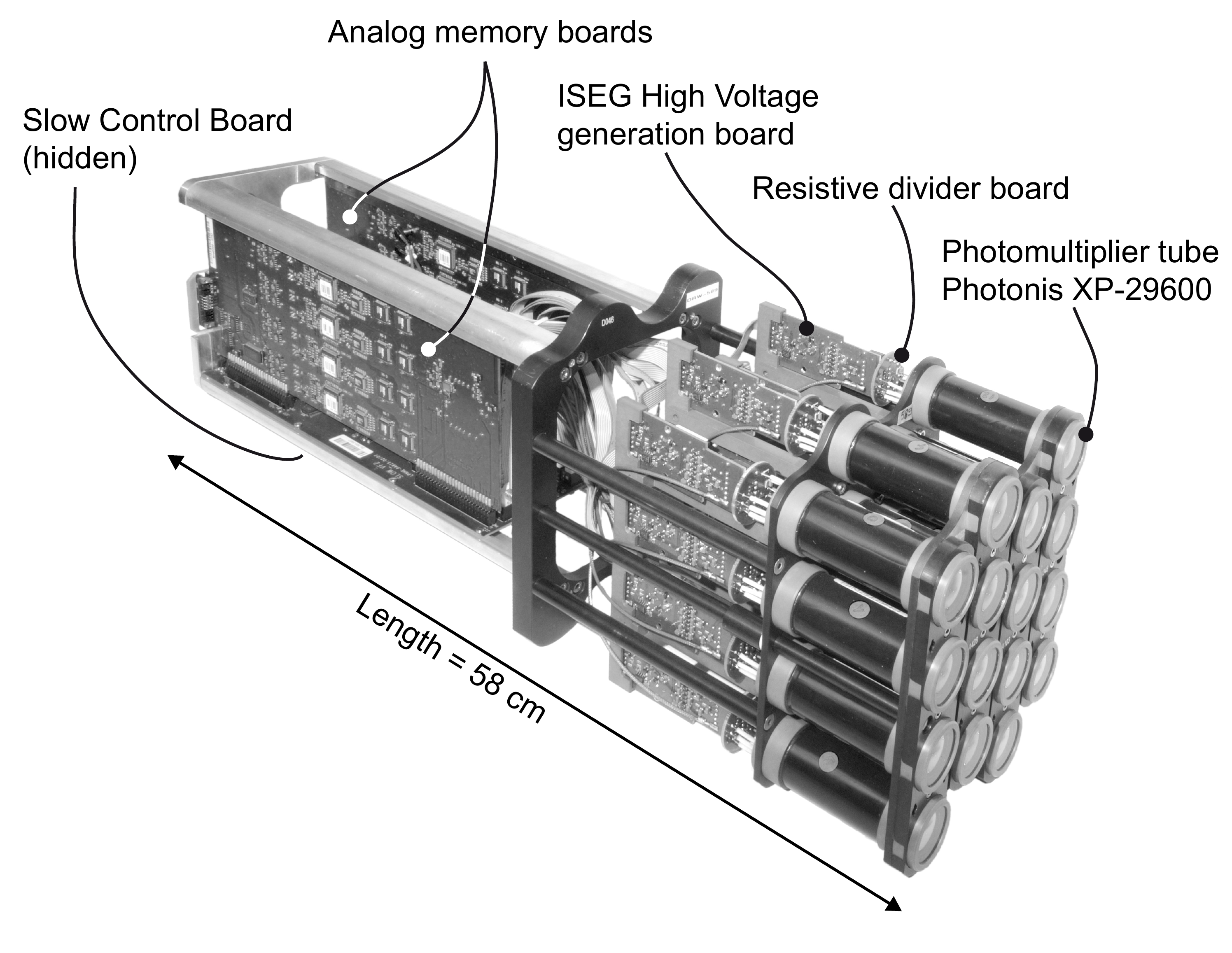} 
   \caption{View of a 16-PMT module of the CT5 camera.}
   \label{fig:photodrawer}
\end{figure}

\begin{table*}[htbp]
   \caption{Comparison of the main characteristics of CT1--4 and CT5 cameras.\label{tab:h1h2comp}}
   \centering
   \footnotesize{
   \begin{tabular}{r|c c} 
      \hline
      \hline
      \textbf{Telescope} & CT1--4 & CT5 \\
      \hline
      \textbf{Mechanics} & & \\
      Total Weight & $\sim$1 tonne & $\sim$3 tonnes \\
      Dimensions (W$\times$H$\times$D) & 160$\times$160$\times$150 cm$^3$  & 227$\times$240$\times$184 cm$^3$\\
      Camera FoV & $5.0^\circ$ & $3.2^\circ$ \\
      Pixel FoV & $0.16^\circ$ & $0.067^\circ$ \\
      Number of drawers & 60 & 128 \\
      Number of pixels & 960 & 2048 \\
      \hline
      \textbf{Photomultiplier Tube} & Photonis XP-2960 & Photonis XP-29600 \\
      PMT diameter & \multicolumn{2}{c}{1'1/8", 29 mm} \\
      Photo-cathode & \multicolumn{2}{c}{Bi-alkali} \\
      Number of dynodes & \multicolumn{2}{c}{8} \\
      Spectral range & \multicolumn{2}{c}{270--650 nm} \\
      Wavelength of peak efficiency & \multicolumn{2}{c}{420 nm} \\
      Anode pulse duration (FWHM) & \multicolumn{2}{c}{2.9 ns} \\
      Peak quantum efficiency (typical) & 25\% & 30\% \\
      After-pulse rate (typical, $>$ 4 p.e.) & \multicolumn{2}{c}{0.05\%} \\
      Nominal gain in H.E.S.S. & \multicolumn{2}{c}{$2\times 10^5$} \\
      \hline
      \textbf{Electronics} & & \\
      Sampling frequency & \multicolumn{2}{c}{1 GSPs} \\
      Memory depth & 128 cells & 256 cells \\
      Analogue bandwidth & 80 MHz & 300 MHz \\
      Nominal width of the integration window$^a$ & \multicolumn{2}{c}{16 ns} \\
      $\Delta_\mathrm{SPE-PED}$$^{b,c}$   & 80 (1 p.e.)    & 50 (1 p.e.) \\
      $\sigma_\mathrm{PED}$$^{b,d}$     & 16 (0.2 p.e.)  & 10 (0.2 p.e.) \\
      $\sigma_\mathrm{SPE}$$^{b,e}$          & 32 (0.40 p.e.) & 19 (0.38 p.e.) \\
      Analogue memory read-out time for 1 event & $\sim$280 $\mu$s & $\sim$2 $\mu$s \\
      Dead-time & $\sim$460 $\mu$s & 15 $\mu$s \\
      Event size per telescope$^f$ & 2.4 kbyte & 7.4 kbyte \\
      Trigger rate$^g$ & $\sim$600 Hz & $\sim$5000 Hz \\
      Power consumption & $\sim$5 kW & $\sim$8 kW \\
      Cooling & \multicolumn{2}{c}{Convection and forced air flow} \\
      \hline
      \hline
   \end{tabular}
   
   $^a${The width of the integration window is programmable.}
   $^b${Average value expressed in ADC counts and obtained for all tested PMTs with nominal gain in the high gain channel, using the fit procedure illustrated in Fig.~\ref{fig:spe}. Values in p.e. are also given in parentheses.}
   $^c${Difference between the mean of the SPE fitted Gaussian and the position of the pedestal.}
   $^d${Pedestal standard deviation.}
   $^e${SPE standard deviation.}
   $^f${Charge mode (see \S\ref{subsubsec:alb}).}
   $^g${Maximum L1 (for CT1--4) and L2A (for CT5) trigger rate.}
}
\end{table*}

\section{Electronics}
\label{sec:elec}

\begin{figure*}[t!] 
   \centering
   \includegraphics[width=7in]{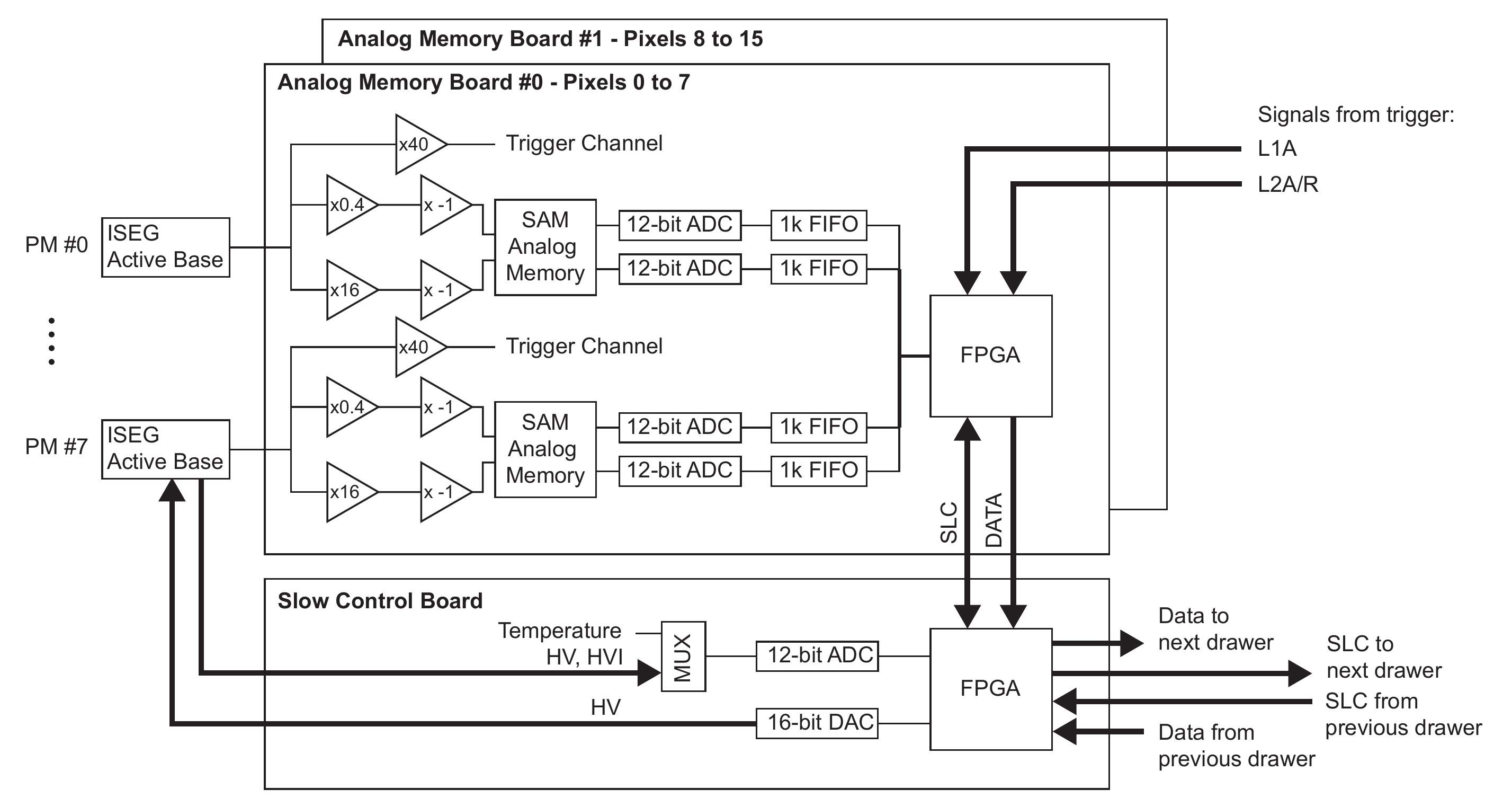} 
   \caption{Schematic of a drawer. Only two physical channels out of sixteen are represented. See the text for a complete description. The trigger channels, as well as trigger signals L1A and L2A/R are represented for completeness. The L1A signal starts the readout of SAM analogue memories and the filling of the FIFOs, while the L2A signal starts the readout of the data from the FIFOs. The detail of the trigger architecture is given in Figure~\ref{fig:trigger}.}
   \label{fig:drawer}
\end{figure*}

\subsection{General description}
\label{subsec:archi}

The front-end of the camera is modular, made-up of 128 identical electronics modules (hereafter called \textit{drawers}, Fig.~\ref{fig:photodrawer}) inserted in the camera from the front so that they can easily be replaced. Each drawer carries 16 PMTs (for a total of 2048 pixels) and three electronics boards: two boards for data acquisition (level 0 trigger (L0), amplification, sampling, conversion of PMT signals) and one board for slow control.

The drawers can receive and send data through dedicated communication buses, using a token-passing mechanism. One bus (parallel, $\sim266\,\rm{Mb/s}$) is devoted to data transfer and the other to slow control messages (serial, $\sim10\,\rm{Mb/s}$). There are 16 data buses of 8 drawers each, and 8 slow control buses of 16 drawers each.

At the rear of the camera, three Compact PCI\,\footnote{\textit{Peripheral Component Interconnect}.} (cPCI) crates are used for data management, trigger, safety and slow control. Each of these tasks is managed by a CPU\,\footnote{\textit{Central Processing Unit}.}, except for data management for which two CPUs are used to optimize data transfer and sustain the required trigger rate of $5\,\rm{kHz}$. Other equipment include five power supplies, a GPS\,\footnote{\textit{Global Positioning System}.} receiver used for time stamping, a terminal server used to access the consoles of the CPUs for maintenance as well as the GPS serial interface. All the CPUs, the power supplies and the terminal server are connected to each other and to the computers on the ground through a Gigabit ethernet switch. 

While normally located in the focal plane during observations, the camera can be unloaded from the telescope and moved inside a shelter during full-moon period or for maintenance and calibration operations. A dedicated, fully-automatic loading/unloading system was designed for this purpose. As a consequence, the number of connections going from the camera to the ground has been minimized to a single $380\,\rm{V}$ power cable, 3 optical fibres for data transfer and trigger, and a compressed air tube. Electrical and optical links are grouped in a single connector.

\begin{sidewaysfigure*} 
   \centering
   \includegraphics[width=8in]{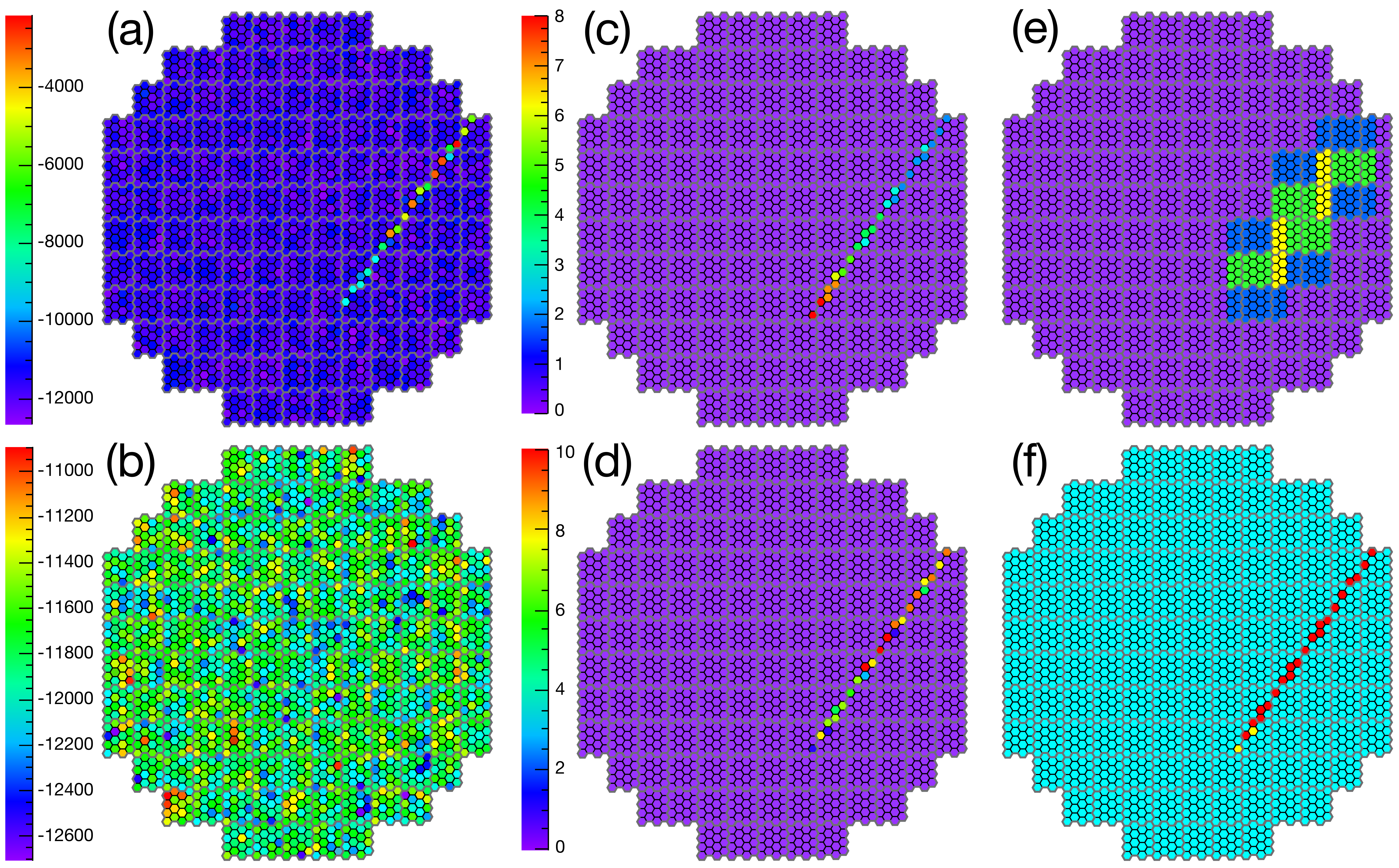} 
   \caption{Event display showing a muon track recorded by the CT5 camera during a test run in October 2011, when the camera was still in Paris, France. (a) Charge distribution in the high gain channel (ADC counts). At the bottom end of the track, the deposited charge decreases significantly. This is due to the fact that the muon is progressively going out of the PMT plane. (b) Charge distribution for the low gain channel (ADC counts). No signal can be seen. (c) $\mathrm{T_0}$ distribution (ns). A gradient from the top to the bottom can easily be seen, showing that the muon was going downwards. (d) Time over threshold (ToT) distribution (ns). As expected for a muon, which deposits a large charge in each PMT hit, this pattern is identical to the charge pattern. (e) L1 sector pattern. Pixels shown in blue, green and yellow participate in 1, 2 and 3 sectors respectively. (f) Ternary image showing the pre-L2 pattern. The cyan colour corresponds to 0, yellow to 1 and red to 2. See the text for details.}
   \label{fig:display}
\end{sidewaysfigure*}

\subsection{Front-end electronics}
\label{subsec:fe}

The front-end electronics consists of three boards in each drawer (Figure~\ref{fig:drawer}). Two boards (analogue memory boards, ``ANALO'') are used to process the signals from the PMTs (8 PMTs per board) and one board (slow control board, ``SLOW'') is used for the control and monitoring of the two ANALO boards connected to it.


\subsubsection{The SAM chip}
\label{subsubsec:sam}

The SAM\,\footnote{\textit{Swift Analogue Memory}} chip is a two-channel analogue memory, based on switched capacitors, that is used as a buffer, in which the photomultiplier tube signal is sampled and stored while waiting for a decision from the fixed-latency trigger electronics. This chip, described in more detail in \cite{SAM}, has been specially designed for CT5 to overcome the main limitations of the ARS chip \cite{ARS} equipping the cameras of the first H.E.S.S. telescopes. SAM can operate with sampling frequencies in the range from 0.5 to $3\,\rm{GSPs}$ (Giga-Samples per second) with a dynamic range exceeding 11 bits. This high signal over noise ratio is obtained thanks to a full differential architecture rejecting the pick-up noise coming from the digital activity inside the chip and on the front-end boards. The chip memory depth has been extended from 128 to 256 cells to be compatible with trigger latencies higher than $100\,\rm{ns}$ for a $2\,\rm{GSPs}$ sampling frequency. The chip analogue bandwidth, increased from 
$80\,\rm{MHz}$ to $300\,\rm{MHz}$,
is fast enough to avoid the widening of the fast pulses from the photodetectors. It allows for use of short integration windows to calculate the pulse's charge, minimizing the effect of the Night Sky Background (NSB). An SPI slow-control link permits to program many parameters and to access to the various modes of operation of the chip.   For each event, only few samples, corresponding to a region of interest within a window located at a fixed time (programmable using a slow-control parameter) with respect to the trigger, are read out. The consecutive samples from the two channels are multiplexed in parallel toward an external 10 MSPs 12-bit 2-channel external ADC for digitization. For a 16-sample event, the readout time is only $2\,\mu\rm{s}$, instead of $270\,\mu\rm{s}$ with the ARS chip. This major improvement allows the camera dead time to be decreased below 1\% for a typical trigger rate of $5\,\rm{kHz}$.

Because of its very good and stable cell-to-cell uniformity of pedestal and gain, the SAM chip only requires a very simple calibration. This operation consists of equalizing 32~DC values using on-chip SPI digital to analogue converters to cancel offsets due to the matrix chip structure. Once this operation has been performed, it remains valid (less than 10\% change of noise) for a few days and for temperature variations within a range of $\pm15^\circ\rm{C}$, with the digitized SAM output data being directly usable for online processing by the FPGA of the analogue memory board, without additional correction.

A batch of 6000~SAM chips has been produced using a $0.35\,\mu\rm{m}$ CMOS process from AMS\@.  Each chip has been tested using a fully-automated procedure to ensure its functionality and compatibility with CT5 requirements. This procedure includes power consumption, baseline and noise measurements but also tests performed using input pulses simulating the photomultiplier response such as signal shape, linearity, jitter and crosstalk characterizations. The on-chip time delay between the trigger and the signal paths defining the timing precision for the beginning of the region-of-interest was also extracted. The uniformity of this parameter is better than $1\,\rm{ns}$ over the whole batch. 
All the parameters extracted during the tests were recorded in a database for further chip sorting and analysis. To guarantee its stability, the test bench was monitored over the 6 months of tests by retesting a set of reference chips every day.  Fewer than 10\% of the chips were found to be out of specifications.


\subsubsection{Analogue memory boards}
\label{subsubsec:alb}

The negative pulse from a PMT is amplified (or attenuated) with three different gains: a gain $\times16$ (high gain) used for low energy events (typically under 400 p.e.\,\footnote{photo-electron.}) and to measure the single p.e. response for PMT calibration, a gain $\times0.4$ (low gain) used for high energy events (up to 5000 p.e.) and a gain $\times40$, used for the trigger signal generation. In addition, the signal is inverted in high gain and low gain channels.

\begin{figure}[t!] 
   \centering
   \includegraphics[width=3.5in]{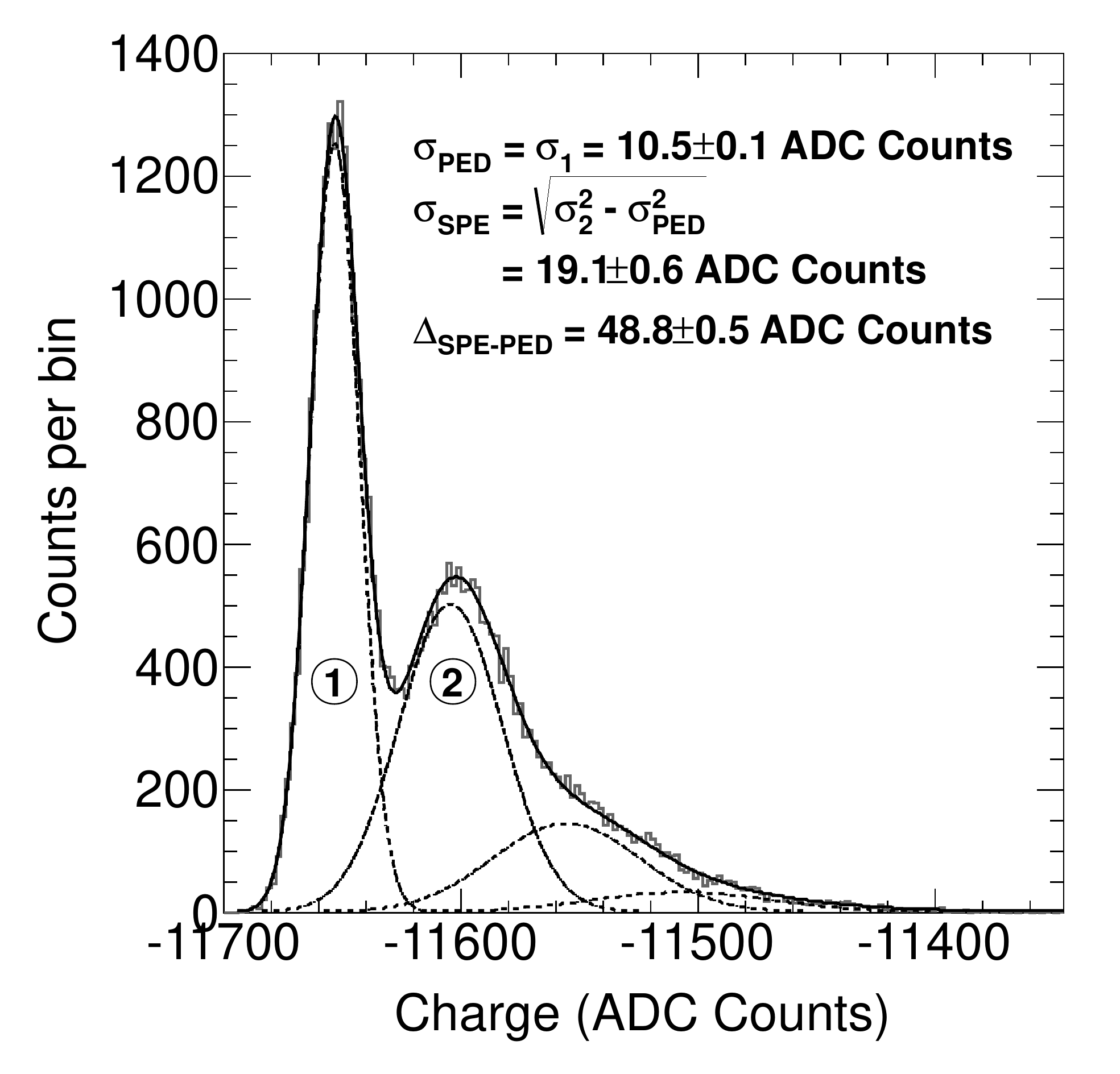} 
   \caption{Typical charge distribution obtained at single photo-electron level with a \mbox{XP-29600} PMT in the high gain channel. The distribution (grey histogram) is fitted with a sum of Gaussian curves (black solid line) with Poisson-distributed relative weights \cite{Bellamy}. The individual components are shown with dotted lines and correspond from left to right to the pedestal (peak 1), 1~p.e. (peak 2), 2~p.e. and 3~p.e. charges respectively. The most important results of the fit are also given: pedestal and single p.e. distribution widths (standard deviations $\sigma_\mathrm{PED}$ and $\sigma_\mathrm{SPE}$), and the separation between the two peaks ($\Delta_\mathrm{SPE-PED}$).}
   \label{fig:spe}
\end{figure}

High-gain and low-gain signals (Fig.~\ref{fig:display}a, \ref{fig:display}b) are stored into two 256-cell analogue memories (within the SAM chip, \S\ref{subsubsec:sam}), acting as circular buffers. Each cell corresponds to a $1\,\rm{ns}$ sample of the signal. If no trigger occurs, the analogue memory content is overwritten as new data are recorded. If a trigger occurs, the buffering is stopped and a given number of cells in the memory is read and digitized. This number is programmable and defines the width of the readout window, which depends only on the PMT pulse width and NSB\@. For all H.E.S.S. cameras, it is set to 16~cells, corresponding to $16\,\rm{ns}$. The digitization is done by a 12~bit, $20\,\rm{MHz}$ ADC\,\footnote{Analogue to Digital Converter. The input range is 0--$2\,\rm{V}$ and the output digital is in the range $\pm$2048 (decimal).}. In order to use the full dynamic range of the ADC, the digitized signal (positive pulse) is shifted towards negative values\,\footnote{This explains why the charge appears to be negative on Fig.
~\ref{fig:spe}, Fig.~\ref{fig:display}a and Fig.~\ref{fig:display}b.}. The 12~bits, corresponding to one $1\,\rm{ns}$ sample of the PMT signal, are then sent to a $1\,\rm{kb}$ first-in first-out (FIFO) chip. This buffer allows to store new events (to a maximum of 64) while waiting for a L2 decision signal (\textit{L2 Accept} or \textit{L2 Reject}, \S\ref{subsec:trigger}) and is therefore extremely important to reduce the overall dead-time.

Each ANALO board has a FPGA chip performing several operations. First, it is used to control the SAM chips. In particular, it stops the buffering and starts the digitization when a L1 trigger occurs. After digitization, the data are stored in the FIFO buffers. Second, the FPGA also controls the readout of the FIFOs and processes their content when a L2 trigger happens. From the 16 signals provided by the PMTs (8 PMTs $\times$ 2 gains), the FPGA builds the data blocks in the mode specified during the configuration of the camera. Two modes can be used: the ``sample mode'', for which the FPGA sends all the samples for each PMT for the two gains, and the ``charge mode'', for which the FPGA integrates the pulse and sends the total charge, as an ADC count coded on 16~bits, for the two gains. In addition, the FPGA computes the time of the maximum of the pulse ($\mathrm{T_0}$, Fig.~\ref{fig:display}c) with respect to the L1A trigger time and the time spent by the signal over a given absolute threshold (ToT, for \textit{time over threshold}, Fig.~\ref{fig:display}d).

Fig.~\ref{fig:spe} shows the charge distribution obtained through the high gain channel when a PMT receives an amount of light which results in the occasional emission of one single photo-electron (SPE). Following \cite{Bellamy} and according to what was done previously in H.E.S.S. \cite{hess1perf}, the distribution is fitted with a sum of Gaussian curves with Poisson-distributed relative weights\,\footnote{Note that this procedure can not be generally applicable to all type of PMTs (see e.g.\cite{Wright}).} to extract the mean of the Gaussian corresponding to the SPE relative to the position of the pedestal (noted $\Delta_\mathrm{SPE-PED}$), the standard deviation of the pedestal ($\sigma_\mathrm{PED}$) and the standard deviation of the SPE ($\sigma_\mathrm{SPE}$). The pedestal variation that can be seen in the figure is mainly due to the PMT and the amplifier and has a standard deviation of 10 ADC counts ($\sim4.9\,\rm{mV}$). The SAM and the ADC contribute only marginally with a noise standard deviation of $\sim730\,\mu\rm{V}$ and $144\,\mu\rm{V}$ respectively.

\begin{figure*}[t!] 
   \centering
   \includegraphics[width=7in]{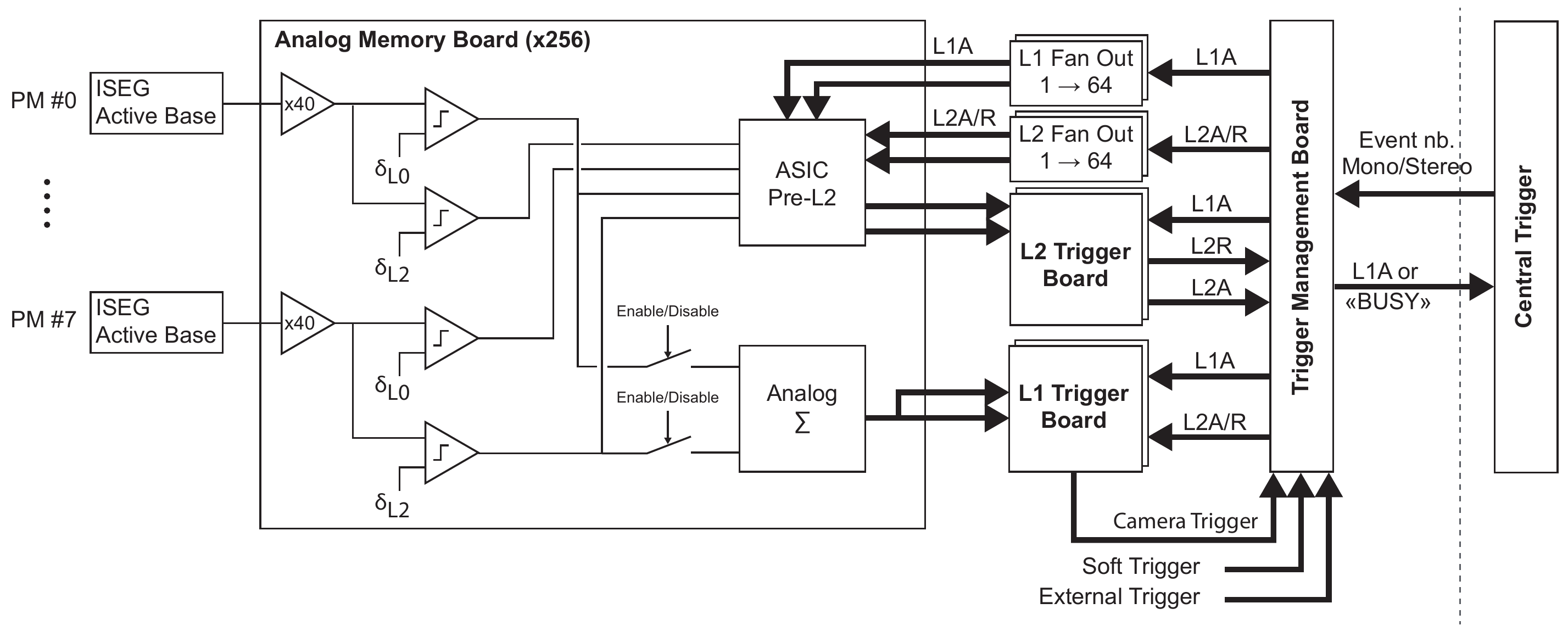} 
   \caption{Block diagram showing the structure of the trigger hardware. Signals are compared to two thresholds $\delta_\mathrm{L0}$ and $\delta_\mathrm{L2}$ for L0 and L2 triggers respectively. Each pixel can be removed from the trigger individually. L2A/R is a pulse for which the width is different for L2A and L2R\@. The dotted line on the right indicates the central trigger is outside the camera, in the control room of the array.}
   \label{fig:trigger}
\end{figure*}

\subsubsection{Slow control board}

The slow control board (SLOW) is used to control all the parameters of the drawer and to encapsulate these parameters for their transmission in the data pipeline. Some quantities are obtained directly by the SLOW board (e.g., the temperature, HV and HVI\,\footnote{Current drawn by the PMT, see \S\ref{sec:conespm}.} for each pixel) and some others through the FPGA located on each ANALO board (e.g., the L0 trigger threshold, the trigger rate for each pixel, $\mathrm{T_0}$ and ToT values for each pixel).

\subsection{Trigger}\label{subsec:trigger}

As in the first four cameras, each pixel can provide a level 0 (L0) trigger and the readout of the whole camera is triggered by a level 1 (L1) decision if several pixels are triggered in a contiguous sector of the focal plane. For the fifth camera, an additional level of triggering was introduced, a level 2 (or L2) topological trigger. The L2 trigger is used to reduce the data rate and select gamma-like events.

During regular data taking, the camera uses the L1 trigger with L2 activated or not, while during calibration runs such as single photo-electron runs, the trigger can be provided by an external source, or generated by the software.

Given the fact the fifth telescope has a lower threshold and a higher event rate than the others, it triggers alone most of the time\,\footnote{The first estimations show that CT5 triggers alone for $\sim$75\% of the events.}.

The cameras of the five telescopes are connected to the central trigger of the array \cite{central} through optical fibres. The central trigger hardware, located in the control-room building, receives trigger signals from the five cameras and provides an event number to the cameras triggered in coincidence. In the case of CT5, the central trigger also informs the camera if it has triggered alone (monoscopic event) or in coincidence with another telescope (stereoscopic event). This information is used by the L2 trigger.

\begin{figure}[t!] 
   \centering
   \includegraphics[width=3.5in]{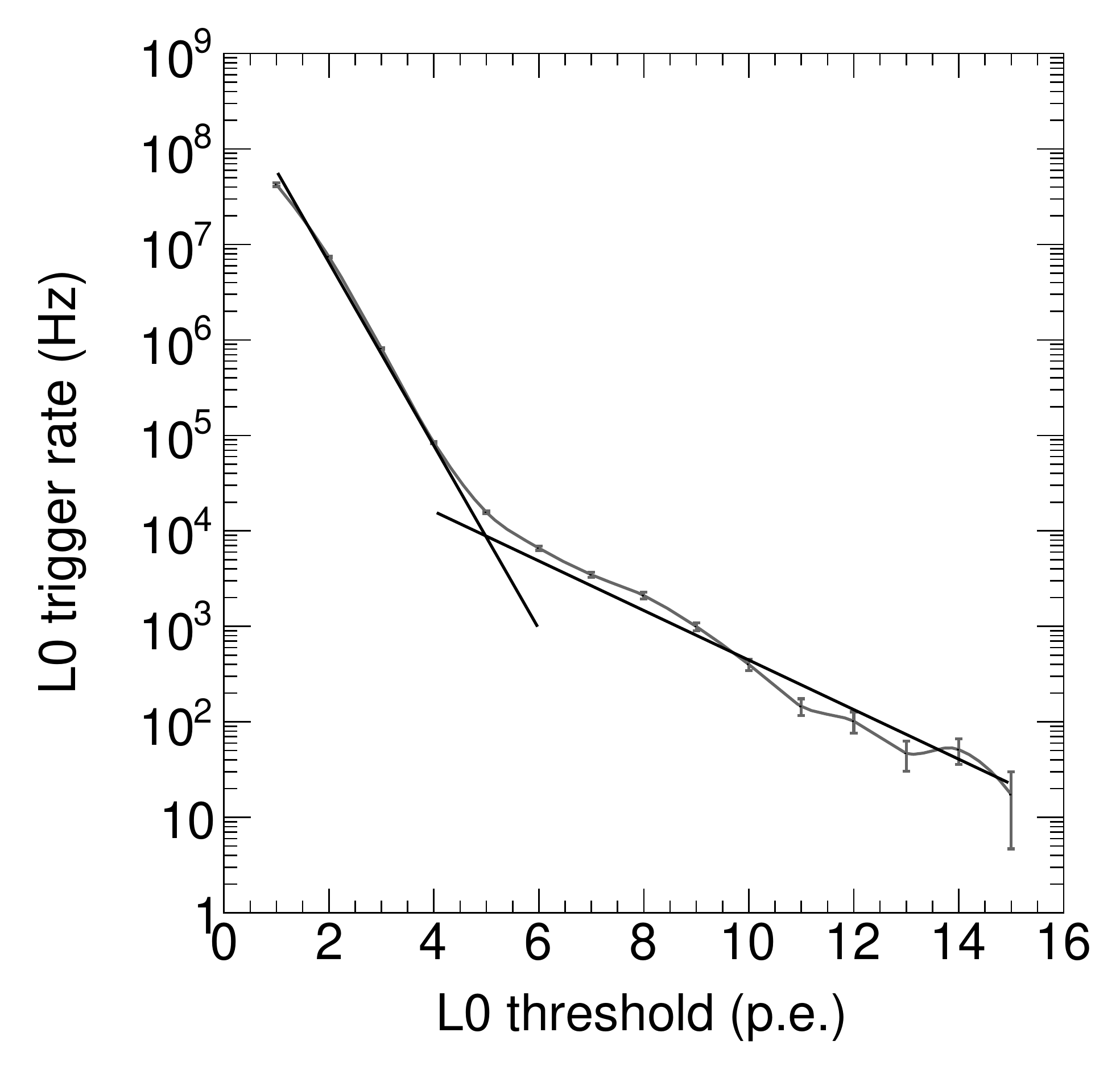} 
   \caption{L0 trigger rate (in Hz) evolution as a function of the L0 threshold (in p.e.) obtained for one of the CT5 PMTs. Each contribution is fitted with a power law. The two functions cross at 5~p.e.}
   \label{fig:afterpulse}
\end{figure}

\subsubsection{L0 and L1 trigger}\label{subsubsec:triggerL0L1}

The L0 trigger is produced on each ANALO board (Fig.~\ref{fig:trigger}). The trigger signal for each pixel is compared to a given threshold $\delta_\mathrm{L0}$ by a comparator. The threshold (expressed in p.e.) is programmable and the same value is set for all the pixels. The outputs of the eight comparators of a ANALO board (one per pixel) are summed and sent to dedicated boards in the trigger crate. It is possible to enable or disable a pixel so that it participates in the trigger decision or not. A pixel is disabled when the HVI exceeds $150\,\mu\rm{A}$. This value is reached typically when a star of magnitude 7--8 enters its field of view.

The value of $\delta_\mathrm{L0}$ is set so as to minimize the influence of the NSB on the trigger rate. To determine $\delta_\mathrm{L0}$, the PMTs are illuminated with a continuous white light, which represents the NSB\@. For H.E.S.S., the NSB rate seen by one pixel can vary from $50\,\rm{MHz}$ to $1\,\rm{GHz}$ (p.e.) depending on the region of the sky observed, for an average of $100\,\rm{MHz}$.

Fig.~\ref{fig:afterpulse} shows the evolution of the L0 trigger rate as a function of the L0 threshold. Two contributions can be seen: below 5 p.e. the NSB dominates, and above this charge the PMT after-pulses  dominate. The L0 threshold is set just below the break of the curve. The present setting is $\delta_\mathrm{L0} = 4\,\rm{p.e}$.

\begin{figure}[t!] 
   \centering
   \includegraphics[width=3in]{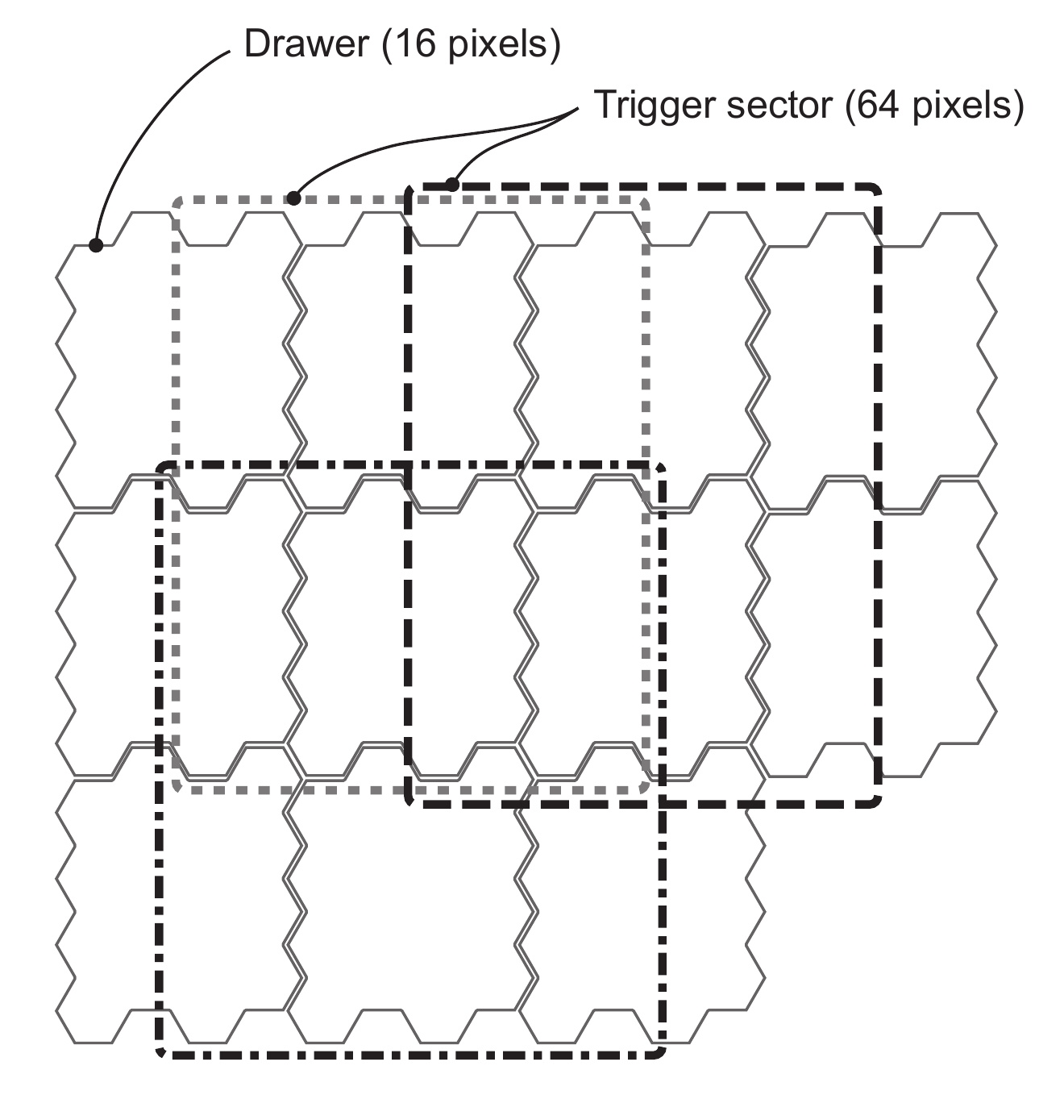} 
   \caption{Schematic of trigger sectors pattern. Eleven drawers are shown, to illustrate the vertical and horizontal overlap pattern of trigger sectors.}
   \label{fig:sectors}
\end{figure}

For the next-level trigger (L1), the drawers are grouped in sectors (Fig.~\ref{fig:sectors}, Fig.~\ref{fig:display}e). A sector is a group of 64 contiguous PMTs, with a vertical and horizontal overlap of 16 and 32 pixels respectively. The L1 trigger is obtained when a sector has at least $N_\mathrm{pix}$ pixels satisfying the L0 condition. $N_\mathrm{pix}$ is currently set to 3.5. The trigger decision is issued by a signal called L1A (\textit{L1 Accept}) in the following. When a L1A occurs, the data stored in SAM memories are read out, digitized, and buffered in the FIFOs. At this point, the SAM chips are available for a new event. Therefore, the use of FIFO buffers in the front-end pipeline allows to release the L1 trigger just 7 to $10\,\mu\rm{s}$ after the L1A\@. This is a major improvement as compared to the value of $460\,\mu\rm{s}$ measured on the CT1--4 cameras, for which no buffering is used. During the readout of SAMs, other triggers can happen, but the camera is not read-out and a ``BUSY'' state is transmitted to 
the central trigger. This information is used to calculate the dead-time of the whole array.

The parameters $\delta_\mathrm{L0}$ and $N_\mathrm{pix}$ are stored in a database and loaded in drawer FPGAs when the camera is configured at the beginning of each night. Different values can be used depending on NSB conditions.

\subsubsection{Level 2 trigger}

As already mentioned, monoscopic events (i.e., events triggered by CT5 only) constitute a large fraction of the L1A rate of the CT5 telescope. Among the latter, only very few are gamma events, so the purpose of the L2 trigger \cite{L2V0,L2V1} is to reject as much as possible of the residual NSB and of the hadronic background.

The 2048 camera pixel signals go through the L0 comparators and a second comparator with threshold $\delta_\mathrm{L2} > \delta_\mathrm{L0}$. The comparator outputs are shaped and stored in a delay-line buffer implemented within a dedicated pre-L2 ASIC in order to compensate the L1 trigger latency. When a L1A occurs, a ternary image\,\footnote{Image in which pixels can take three values, 0, 1 or 2.} giving the list of pixels above $\delta_\mathrm{L2}$ and $\delta_\mathrm{L0}$, resynchronized with respect to the L1A signal, is transmitted to the L2 trigger system (Fig.~\ref{fig:display}f). 

The L2 behaviour depends on the type of the event: monoscopic or stereoscopic. This information is provided by the central trigger. 

In the case of monoscopic events, the ternary pixel map transmitted to the L2 trigger is processed by a real-time implementation of the L2 selection algorithm based on the detection of clusters of active pixels, and the computation of first and second order Hillas moments \cite{Hillas} to identify gamma-like or hadron-like images. In the first case, a signal L2A (\textit{L2 Accept}) is generated, which triggers the read-out of the FIFOs for the whole camera. Otherwise, an L2R signal (\textit{L2 Reject}) is issued to reject the current event: the event is dumped from the FIFOs and discarded, and the L2 system is ready to process a new L1A event. 

In the case of stereoscopic events, the L2 trigger skips the selection algorithm and generates a L2A signal. In other words, a L2A is generated for all events triggering more than one telescope.

When a L2A occurs for an event, all the data corresponding to that event are read from the drawer FIFOs, for all pixels. This operation is performed by the FPGAs and requires $\sim250\,\mu\rm{s}$ in total. The readout is done in parallel for the sixteen data buses. Then, all data are sent to FIFOs located in the data management crate. This second level of buffers is used to store the data blocks until they are merged to form the full event and sent to a farm of computers located in the control-room building, for further processing \cite{daq}.

The overall dead-time of $15\,\mu\rm{s}$ quoted in Table~\ref{tab:h1h2comp} includes the time to release the trigger plus the time interval necessary for the central trigger to send the event number and event type (monoscopic or stereoscopic).

\begin{figure*}[t!] 
   \centering
   \includegraphics[width=7in]{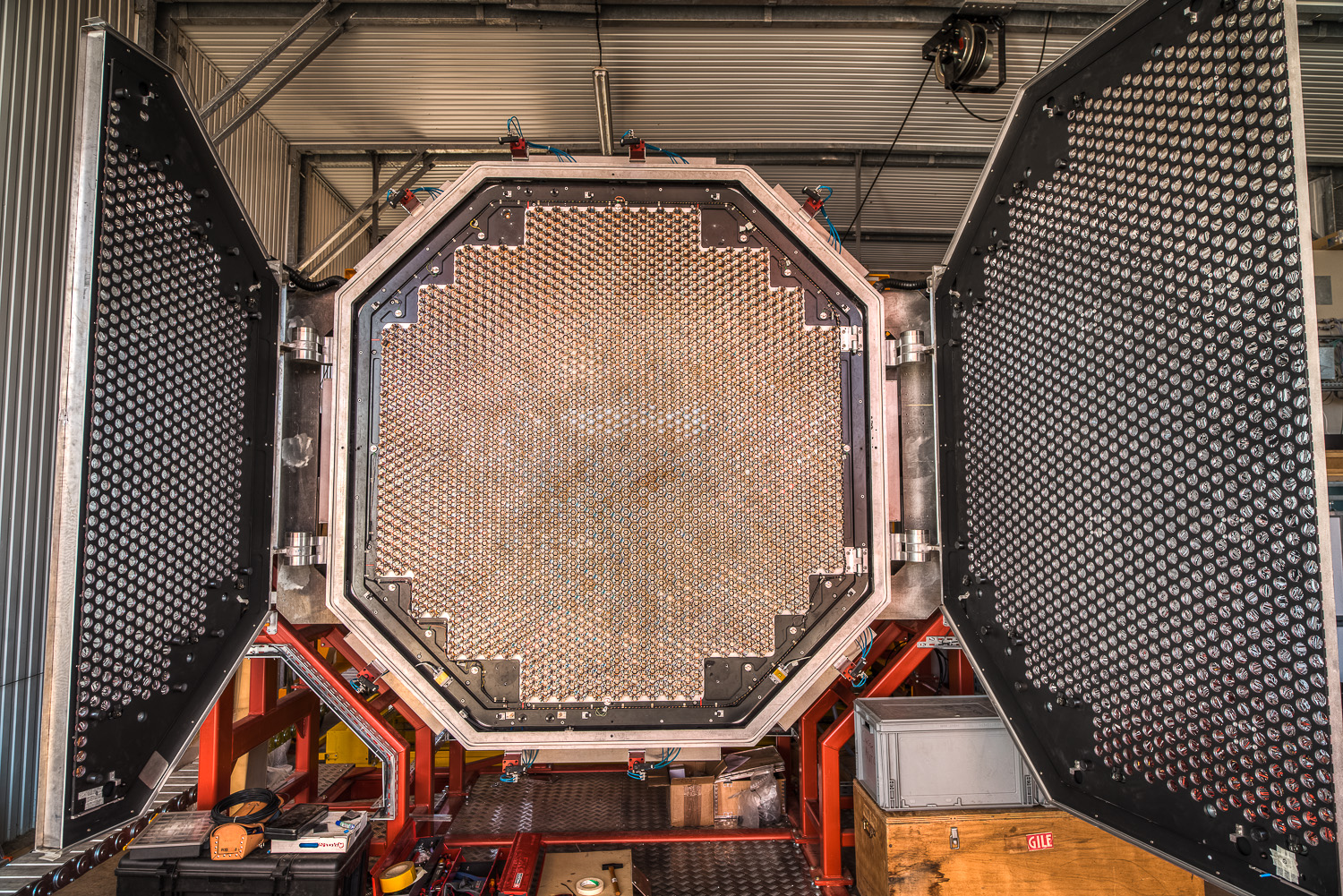} 
   \caption{Front view of the camera with doors open showing the Winston cones and the Mylar plates secured on the doors. Credit: C.~Foehr (MPIK Heidelberg).}
   \label{fig:camerafront}
\end{figure*}

\section{Safety and slow control management}
\label{sec:slc}

The safety and slow control system monitors the camera environment parameters (temperature, level of ambient light) and the status of different camera sub-devices such as the position of the lids, the position of the Mylar plate (see \S\ref{sec:meca} and \S\ref{sec:calib}), etc. It also controls the fans and allows to enable or disable the opening/closing of the lids, the PMT high voltage and the camera power supplies. The outputs of the positioning LEDs\,\footnote{\textit{Light Emitting Diodes}.} (\S\ref{sec:calib}), seen by a CCD\,\footnote{\textit{Charge-Coupled Device}.} camera located on the dish, can be controlled individually. These outputs are used to measure the camera position precisely.

In addition, the system includes safety functions that protect the camera in case a failure in critical components is detected. For this purpose, the system is able to take fast, standalone decisions. For example, the system can make an emergency stop of the PMT power supply, or prohibit the opening of the camera lids if the safety conditions are not fulfilled.

This main logic is implemented in a FPGA hosted in a cPCI crate placed at the rear of the camera. Due to the high number of required interfaces and connections, the safety and slow control system is implemented in several boards. One board is dedicated to the main logic, the cPCI interface with the CPU crate and the photo-diode reading. A second board drives the positioning LEDs. Three others are dedicated to the fan control and the last one to the camera temperature monitoring.

Even though the safety and slow control hardware logic is located inside the camera and interfaces directly with the different devices, the system is piloted remotely from the H.E.S.S. array data acquisition system \cite{daq} through a CPU board. Thus, the slow control can be configured with different levels of safety. In addition, all monitored data or the standalone decisions taken locally can be reported. As soon as the camera is powered, the safety system is active.

\section{Mechanics}
\label{sec:meca}

\subsection{Introduction: Requirements}

The mechanics of the CT5~camera relies on the same general concepts successfully applied in the previous cameras.
Because of the much larger size of this camera relative to those of CT1--4, slight adjustments of the camera concept have been made and the experience gained in building the CT1--4 cameras allowed us to improve the mechanical design in some respects. The requirements of the CT5 camera are given in Table~\ref{tab:h1h2comp}, with a special care on the minimization of the shadowing by the camera, improved reliability of the electronics cooling and improved accuracy of the pixel location. 

The camera is attached to the telescope quadrupod such that its distance to the dish can be remotely controlled to adjust the image plane to focus on the shower maximum (\S\ref{subsec:autofocus}). It can also be unloaded and loaded remotely by a trolley system to allow it to be parked into a shelter for maintenance and protection. As a consequence, the camera does not need to be fully waterproof.

The fully equipped camera has a total weight of 3 tonnes, which corresponds to the weight budget allowed by the design of the mechanical structure of the telescope.

\begin{figure}
\begin{center}
  \includegraphics[width=3.5in]{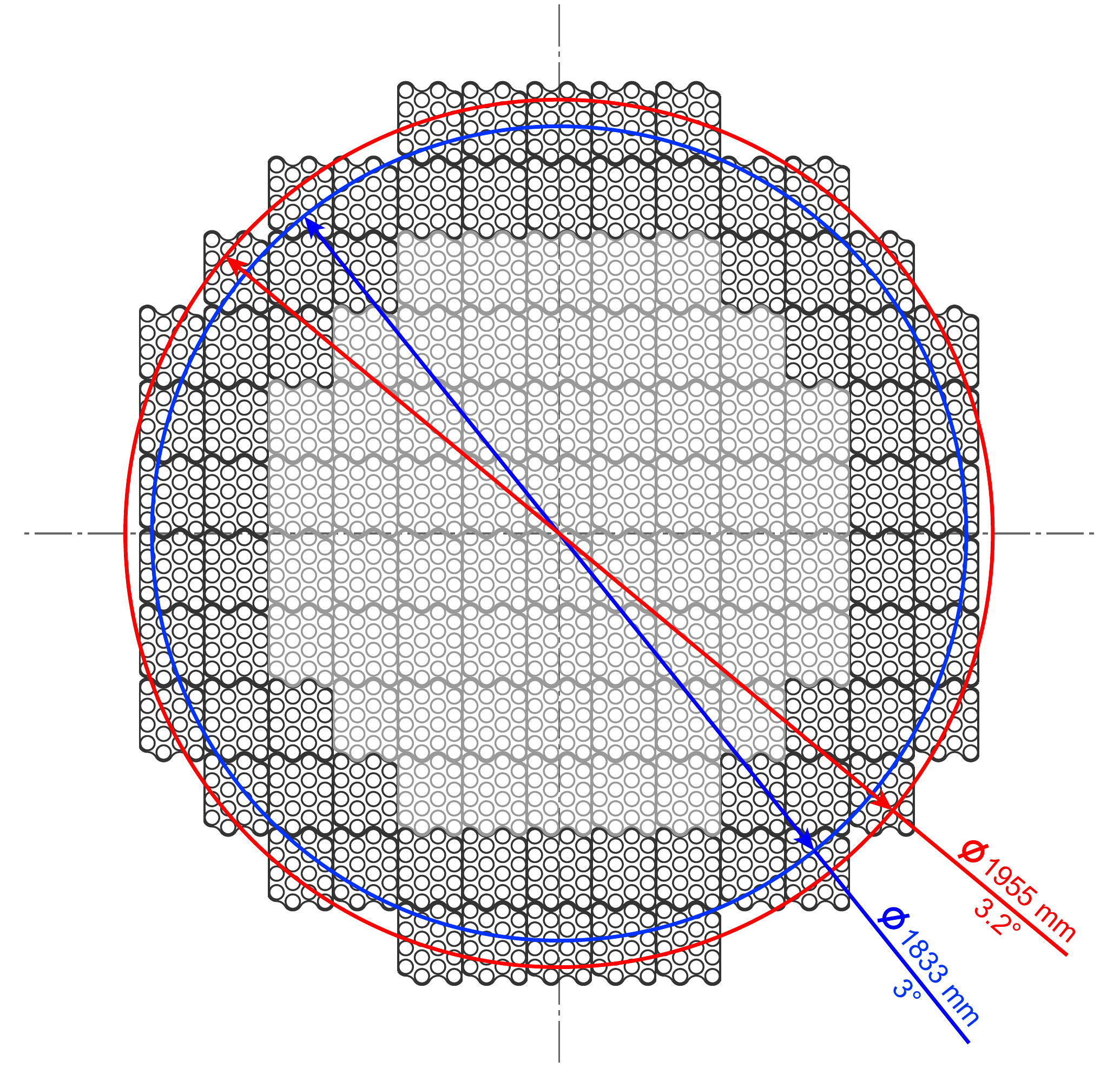}
\end{center}
\caption{Schematics of the pixel layout of the focal plane. The diameter of the focal plane fully covered by pixels is $3^\circ$ and its outside diameter with a partial coverage is $3.2^\circ$. For comparison, the lighter grey area in the centre corresponds to the CT1--4 camera layout (same physical scale).}
\label{fig:FP}
\end{figure}

\subsection{Focal Plane}

The heart of the detection system of the camera is located close to  the focal plane (Fig.~\ref{fig:camerafront}) and hosts the photo-detection system composed of 2048~pixels arranged as shown in Fig.~\ref{fig:FP} and consisting of the Winston cones, the PMTs (\S\ref{sec:conespm}), and their associated electronics, i.e., the front-end boards (\S\ref{subsec:fe}).

The PMTs and the front-end boards are embedded into a modular mechanics structure called a \textit{drawer} (Fig.~\ref{fig:photodrawer}). They can be manually extracted from the camera front, which requires to unscrew just two screws. This system, very similar to the drawer concept used for CT1--4, brings a high modularity for the assembly, the integration, the use of interchangeable drawers and the maintainability. The power and electronics connections with the rest of the camera are made via three connectors placed on the rear of the drawers. These connections impose that the drawers must be positioned with an accuracy better than $0.5\,\rm{mm}$.

\begin{figure}[t!]
\begin{center}
  \includegraphics[width=3.5in]{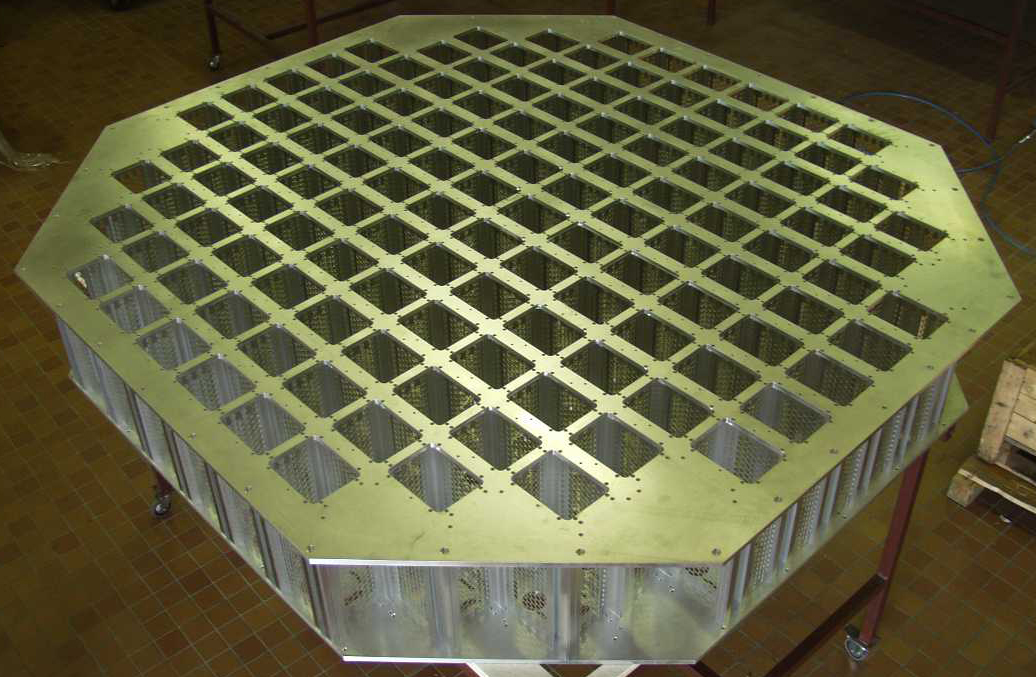}
\end{center}
\caption{Picture of the sandwich.}
\label{fig:Sand}
\end{figure}

To form the focal plane and to insure a good electrical connection when the drawers are inserted from the front, the drawers are placed into the central part of the mechanics, called the ``sandwich'' (Fig.~\ref{fig:Sand}). It is made of two thin aluminium sheets ($2151\times 2010 \times12\,\rm{mm}^3$) assembled from 146~aluminium profiles (with a length of $326\,\rm{mm}$), developed specifically for this camera. With this system, the drawers' location within the focal plane has an accuracy of $0.1\,\rm{mm}$. By design, this sandwich's second function is to provide mechanical stiffness to the whole camera. The management of all the interfaces (mechanical, electrical, optical) was thus an important aspect during the design process.

The system of Winston cones is placed in front of the sandwich and drawers. They are held and placed within the telescope optical focal plane by a large aluminium plate ($1990\times 2090\times 27\,\rm{mm}^3$), on which the Winston cones are clipped with plastic washers. As the cone localisation determines the astrometric accuracy of the pixels and the optical transmission between them and the PMTs, this aluminium plate is machined with a high accuracy such that the Winston cones are localised at $\pm0.2\,\rm{mm}$ on it (or 0.01 arcsec). Its distance to the drawers ($1\,\rm{mm}$) is finely controlled by screws distributed over the whole surface of the plate and limiting the effect of its gravity deformation. These accuracy values have been measured by a company specialised in metrology\,\footnote{E.S.P.A.C.E. S.A., based in Saint-Andr\'e-Les-Eaux in France.}.

\begin{figure*}[ht!]
\centering
\includegraphics[width=7in]{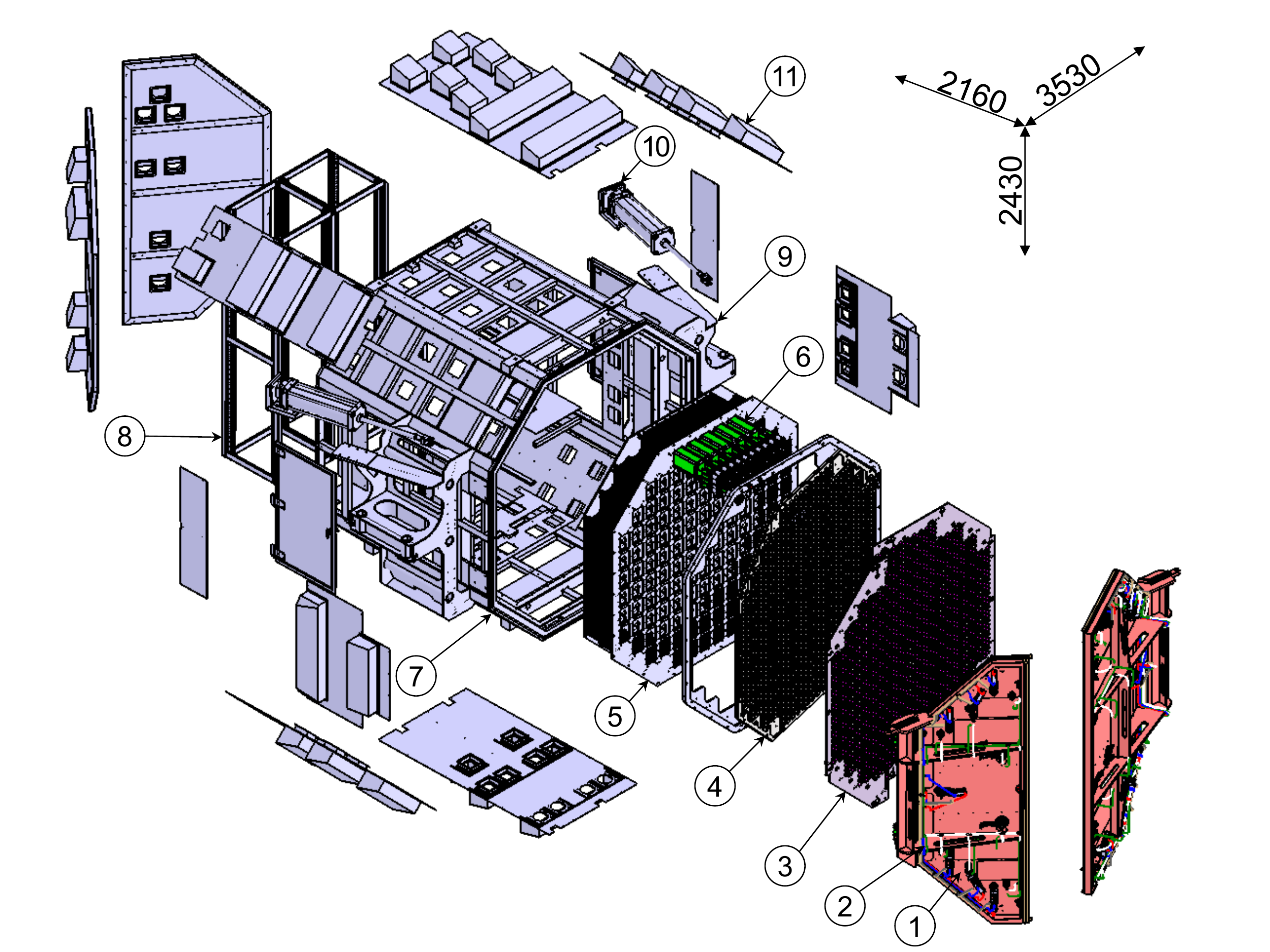}
\caption{Exploded view of the camera with all its components. 1: Automated front lids. 2: Pneumatic systems for the calibration devices. 3: Mylar plates. 4: Winston cone plate. 5: Sandwich. 6: Drawers. 7: Camera body. 8: Electronics racks. 9: Camera attachment interfaces. 10: Pneumatic pistons. 11: Camera envelope.}
\label{fig:ExplodedView}
\vspace{-0.5cm}
\end{figure*}

\subsection{Camera Body}

The instrumentation of the focal plane and all the camera equipment and services are integrated into the camera body (Fig.~\ref{fig:ExplodedView}). This is a skeleton of thin aluminium profiles, whose shape has been specially designed for the project. The profiles are welded together to avoid any ageing. A company\,\footnote{SOPRANZI S.A.S., based near Lyon in France.} specialised in producing and integrating large and complex aluminium pieces assembled these profiles to the required accuracy, reaching a measured $\pm 1\,\rm{mm}$ of the external dimension of the camera body.

The camera envelope, made with thin aluminium plates, is riveted on this skeleton. The plates provide a thermal isolation from the direct sun-light power. This envelope integrates also a system for cooling, as described in~\S\ref{subsec:cooling}.

This skeleton holds also all the internal equipment, the sandwich with its 128~drawers, the two 19-inch electronics racks carrying electronics boards and services such as the power supplies, and the front, lateral and rear lids. It should be noted that front lids have several functions. First, they protect the focal plane from the outside environment and second, they hold calibration instrumentation (\S\ref{sec:calib}). They are motorized by pneumatic pistons and controlled either manually or remotely by the camera Slow Control system (\S\ref{sec:slc}). The pressurized air is generated by an on-ground air compressor and is brought to the camera by pneumatic tubes fixed on the telescope structure. Three air-tanks are used along the path to create buffer volumes for safety\,\footnote{One on the compressor on the ground, one in the telescope and one in the camera.}. The choice of a pressurized air system fully managed by pneumatic logic has been driven by the need to be able to close the camera lids automatically in case of a general power failure.

Finally, the skeleton supports the two large mechanics parts making the interface to the telescope quadrupod structure (\S\ref{subsec:autofocus}).

During the design process, finite element simulations (FEM using the ANSYS module within the CATIA software \cite{catia}) have been carried out by modelling mechanics pieces as hollow components. This allowed to optimise the design with a focus on the ratio of stiffness to weight and to achieve a rigid and accurate body. This process led us to choose components made of aluminium profiles welded together, despite the fact that only a few companies have the technology and the know-how to produce such large objects. The main camera dimensions are summarized in Table~\ref{tab:h1h2comp}.

\subsection{Ventilation system for the electronics}
\label{subsec:cooling}

The electronics components dissipate about $8\,\rm{kW}$ of thermal power inside the camera. For proper operation and extended lifetime of the electronics, this heat must be evacuated so that the temperature of the camera is kept as stable as possible during data taking. 

As for the CT1--4 cameras, the cooling is obtained by convection and by forced-air flows. Fresh filtered air is injected inside the camera by large fans located on the camera envelope and protected from rain by lightweight hoods. This fresh air is convected to the rear of the sandwich. For each drawer, two fans blow the air towards the electronic boards. Then the warm air is drawn by large fans that blow it out of the camera. The speed of the different groups of fans can be changed under slow control, depending on atmospheric conditions.

This ventilation system, despite its simplicity, allows to efficiently cool the FE electronics and the instrumentation located into the racks at the rear of the camera.

\subsection{The focusing system}
\label{subsec:autofocus}

The nominal distance between the position of the camera and the centre of the mirror is $36\,\rm{m}$, which corresponds to the focal length of the telescope. However, as the atmospheric showers develop at a finite altitude between $10\,\rm{km}$ and $30\,\rm{km}$, the position of the camera can be adjusted to optimize the trigger rate and angular resolution depending on observation conditions (in particular the zenith angle) \cite{razmik}. The change in position is done thanks to a system which can move the camera along the optical axis of the telescope. This focusing system has been designed to operate in the sometimes harsh environmental conditions of the Namibian site. Its main technical specifications are given in Table~\ref{tablefocus}.

\begin{table}[t]
\begin{center}
\caption{\label{tablefocus} Main requirements of the focusing system.}
\begin{tabular}{r|c}
\hline
\hline
Focus range & 230 mm from 35.93 to 36.16~m  \\
Depth of field & 7.6~km to $\infty$ \\
Position accuracy & 0.35 mm \\
Moving speed & $0.015\ \rm{m\,s}^{-1}$ \\
Zenith angle & $+125^{\circ}$ to $-90^{\circ}$ \\
Temperature range & $0^{\circ}\rm{C}$ to $50^{\circ}\rm{C}$ \\
\hline
\hline
\end{tabular}
\end{center}
\end{table}

\begin{figure}[t!]
  \centering
  \includegraphics[width=3.5in]{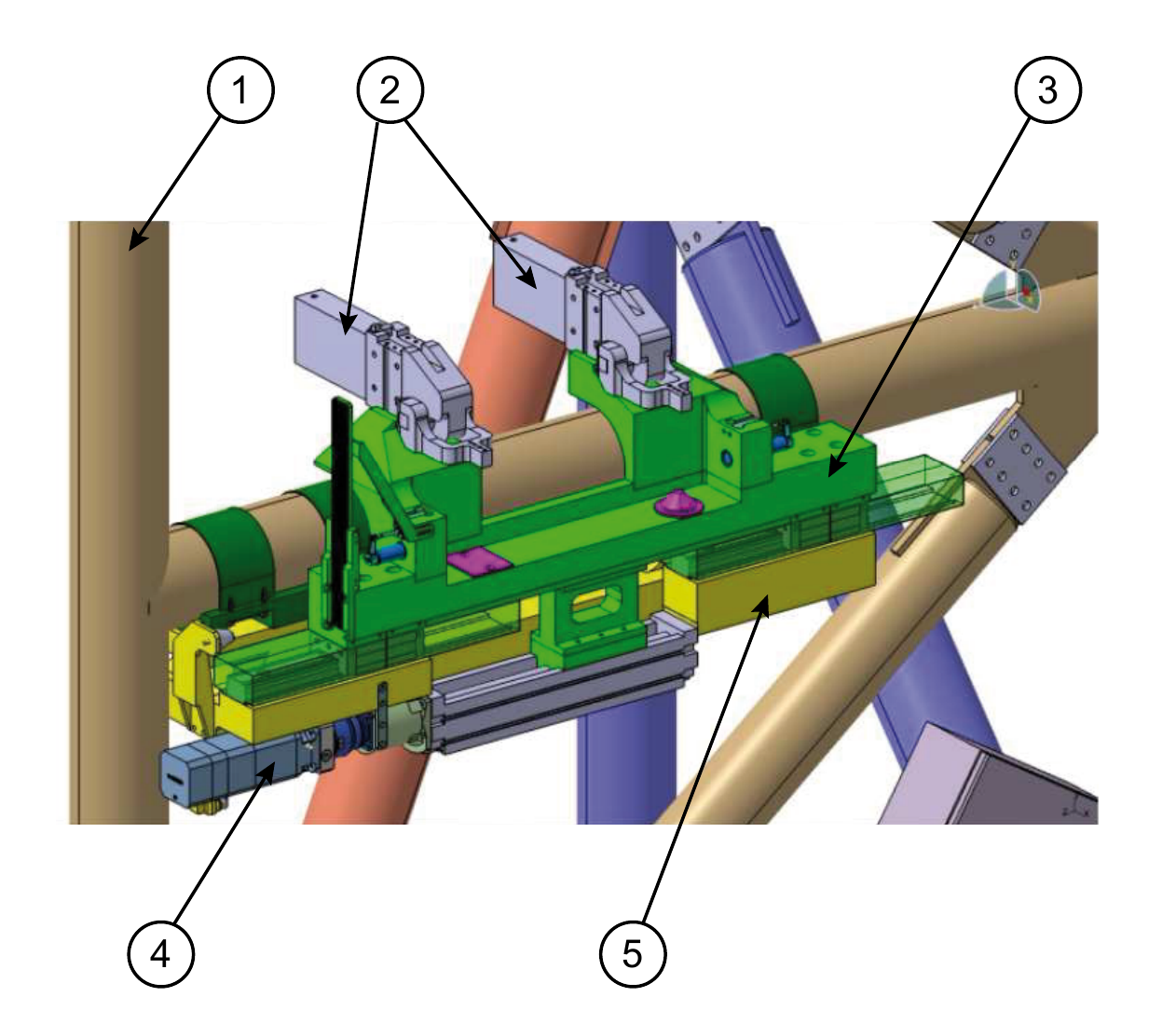}
  \caption{Components of the focusing system. One side of the system is shown here, the other one is symmetric. 1. Structure of the quadrupod. 2. Locks used to keep the camera fixed to the focusing system. 3. Moving part of the focusing system.  4. Motor. 5. Static part, fixed to the structure of the quadrupod.}
  \label{fig:autofocaus}
 \end{figure}

The mechanical design of the focus system aims at two main functionalities: 
\begin{itemize}
\item Install and lock the camera into the telescope quadrupod structure with an accuracy of $0.35\,\rm{mm}$. As mentioned earlier, the camera can be loaded or unloaded from the telescope for maintenance or calibration purposes. The camera is locked and unlocked in the focal plane by using a pneumatic system composed of four toggle fasteners and four jacks (Fig.~\ref{fig:autofocaus}).
\item Adjust the position of the camera along the optical axis, within the camera support mast structures. 
\end{itemize}

The focusing system is made of two components, one is fixed to the telescope structure while the other can move in order to allow the translation of the camera. This motion is driven by two ball screws and two brushless motors which guarantee a speed of $0.015\,\rm{m\,s}^{-1}$ and a position accuracy of $0.05\,\rm{mm}$. The camera can be moved by $230\,\rm{mm}$.

The data acquisition software controls the camera position during the focusing process. While the translation has started, this process has to be guaranteed until the end of the programmed movement, even in case of an unexpected stop (e.g., due to a power cut). The system is controlled through a Programmable Logic Controller architecture composed of two CPUs, two field-buses and two specific variators used to react to power cuts and to manage synchronized operations of the two motors. 
The first estimations, obtained through simulations, have shown that adjusting the focus height to the altitude of the shower maximum allows an increase of 5--10\% of the trigger rate and a 10\% improvement on the angular resolution at low energies ($<80\,\rm{GeV}$), without any alteration of the performances at higher energies.

\section{Calibration Instrumentation}
\label{sec:calib}

To be able to properly analyze and interpret the data produced by the camera, it is necessary to be able to monitor the gain of each channel, the uniformity of the PMT responses, and the exact position of the camera in the focal plane.

The conversion factor between ADC channels and charges is monitored by illuminating the camera such that each PMT's photo-cathode emits a single p.e. (SPE), on average. This calibration can be done either with the camera in the shelter or outside when the camera is in the focal plane. In the first case, a pulser is used to generate flashes of blue light ($\lambda = 340\,\rm{nm}$) at the required intensity and rate, and the light illuminates the PMTs directly through a diffuser. The same pulser is used to trigger the camera via the external trigger input.

When the camera is in the focal plane, it is necessary to filter out the light from the NSB\@. For this, Mylar sheets are placed in front of the Winston cones, providing an attenuation of light intensity by a factor $10^{5}$. As a consequence, to illuminate the PMTs at a SPE level, the intensity of the calibration light source, located at the centre of the telescope dish, has to be far greater than the intensity of the source in the shelter. The Mylar sheets (Fig.~\ref{fig:camerafront}) are held by two large aluminium plates that can either be in contact with the cone plate during SPE runs, or kept attached to the front lids by the use of small pneumatic clamps during normal observations. When in the focal plane, the camera is triggered by an additional PMT, not covered by the Mylar plate, located on the top of the focal plane. When the flasher illuminates this calibration PMT, it delivers a signal that, shaped and delayed, is used as a trigger signal.

The calibration light source is used to provide either SPE pulses or Flat Fielding patterns, used for pixel inter-calibration. A $5\,\mu\rm{J}$ pulsed laser ($532\,\rm{nm}$) is installed in a stainless steel container mounted at the centre of the telescope dish. This container is under constant overpressure to avoid environmental pollution. The repetition rate of the flashes is software selectable and set at $1\,\rm{kHz}$ by default. A holographic filter expands the beam to a flat distribution over the camera FoV\@. A number of neutral density filters, mounted in front of the laser on a motorized filter wheel, are used to vary the intensity of the light output. A dedicated processor board is mounted inside the camera housing and ensures control functions and communications with the DAQ\@. The role of this board is to control the laser, the filter wheel, as well as the delay\,\footnote{This programmable delay is used to synchronize the flash of the laser with the time the camera is triggered.} and the threshold applied to the 
calibration PMT signal to deliver a proper external trigger signal to the camera.

For pixel inter-calibration (the \textit{Flat Fielding}), the camera is illuminated directly by the laser (i.e., the Mylar plates are on the opened front lids) at an intensity of $\sim$200 p.e. The non-uniformity of the light intensity does not exceed 12\% (standard deviation). The charge seen by each PMT is analyzed off-line and correction coefficients are applied to each pixel to compensate for collection and quantum efficiency inhomogeneities.

It is necessary to have the exact position of the camera in the focal plane in order to build an accurate model of the pointing and optical system for use during the data analysis \cite{Pointing}. This has a direct effect on the point spread function of the instrument. 16~red LEDs ($\lambda= 637\,\rm{nm}$), with intensity controlled individually by the slow control system (\S\ref{sec:slc}), are accurately fixed on the Winston cone plate, and are monitored during observations with a dedicated CCD camera located at the centre of the dish. In that case, the camera lids are opened. When taking measurements for the pointing correction model and for the main mirror facets alignment, the camera front lids are closed and used as a screen to project the images of stars. In that case, the light from the LEDs passes through the lids via specific holes which, in other circumstances, are closed by movable covers remotely controlled by pneumatic actuators.

\section{Software}
\label{sec:soft}

From the software viewpoint, the CT5 camera functionally consists of an array of 2048 photomultipliers which take the digital images, and which are serviced by various equipment such as power supplies, front-end electronics, triggering apparatus, gauges, fans, actuators, etc. The control of the whole apparatus must be done remotely through a graphical user interface running on a control computer, and the different types of data generated during a run must be sent back to a central farm of analysis computers.

The embedded software of the CT5 camera interfaces between the camera hardware and the array control room, to which it is connected via a gigabit Ethernet link. It addresses the following: programming and monitoring the various hardware parameters (``slow-control''), acquisition of the shower image data, acquisition of the trigger data, and real-time on-board control of the camera security.

Each of these points is handled by dedicated branches comprising a cPCI crate, a processor board, miscellaneous proprietary interface boards and buses. 

As seen by the software, a branch is a server application running on the branch processor (under a standard Linux system with POSIX real-time extensions) and servicing requests sent by the camera controller software as a client. The server of the slow-control branch is particular because, besides being in charge of controlling the hardware parameters, it is used as a gateway for communications between the control computer and the three other servers.

Each server is organized as a multi-thread process. The scheduling policy of the different threads is set so as to ensure the deterministic behaviour needed for real time applications. In such a scheduling policy, the running thread keeps running indefinitely until either it voluntarily relinquishes the control of the processor or until it is pre-empted by a thread of greater priority. In each server, one thread listens to an Ethernet socket in order to read the commands sent by the controller and a variable number of other threads execute the requested tasks. The thread listening to Ethernet is given the highest priority of execution.

The command and control server has three principal tasks to execute periodically (in addition to several transitory on-request ones): read the GPS time and distribute it to the data and trigger branch servers, so that they can remain correctly synchronized; monitor the front-end electronics voltage, current and temperature, and shut down the power supplies if anything goes wrong; ensure that the on/off status of each photo-multiplier is properly managed throughout the run, according both to the predicted movement of the stars through the detector and to any transient or accidental event that may occur during the programmed observation (e.g., shooting stars).

The server dedicated to image data acquisition has only one important task to do: read quickly the data from the drawers and send them as formatted blocks to the farm of analysis processors via a 1~Gbits/s Ethernet link. In order to avoid overloading the data receiver, several nodes are used and a switch of the receiving node is made every 4 seconds. To be able to sustain high acquisition rates (more than 6~kHz), the server is currently distributed over two separate processors operating alternatively, one reading the data while the other is writing them. This way, constant data reading can be maintained whatever happens on the receiver side. This is a crucial feature to avoid the loss of data coherency that may happen if data overflow occurs at the front-end level. Synchronization of the two processes is carried out through a set of customized ``test-and-set'' devices implemented on the proprietary interface board. The server can automatically reconfigure itself to execute on a single processor if necessary. 
This is possible, however, only if the acquisition rate is relatively low (less than 2~kHz) and if the network is sufficiently stable.

The server dedicated to the acquisition of the ``Trigger'' data is similar in functionality to the data acquisition server but it is structured differently because it always has to run on a single processor. Besides the thread listening to the Ethernet socket, it is composed of two threads instead of one. One low-priority thread reads and formats the data from the trigger front-end boards (hereafter called the ``reading thread'') and the other (the ``sending thread'') sends them with higher priority, packed into multi-events bunches, to the farm of processors. As writing to the socket is done in synchronous mode, it may happen that the sending thread gets stuck while waiting for the end of a data transfer. In that case the reading thread simply goes on reading without trying any longer to transfer the data. This allows to guarantee the events' integrity, even when the trigger rate is very high. Besides its main function of data transfer, the server also executes miscellaneous on-request transitory tasks related to the management of the trigger set-up.

Regarding calibration, the embedded software (running on dedicated cards: one controlling the source unit and the other the trigger unit) has also to implement the interfaces between the calibration hardware on the one hand and the DAQ processes on the other hand, in order to program and monitor the various hardware parameters.

\section{Conclusion}

The most important requirement of the CT5 camera is to be able to handle a trigger rate of several kHz without any data corruption, i.e., a factor of ten higher than the previous H.E.S.S. cameras. This goal was achieved through a complete re-design of the electronics and of the acquisition pipeline. In particular, the dead-time was reduced significantly using new analogue memories, introducing FIFO buffers at several stages of the acquisition chain and optimizing the data transfers.

The CT5 camera brings a lot of improvements, not only for the main acquisition electronics. CT5 is the first Cherenkov telescope in the world to actively move the focal plane in order to adjust to the depth of the shower development in the atmosphere. The camera can be also unloaded from the telescope focal plane to protect it from the weather conditions or for maintenance operations. Many of these features could be considered seriously for a future project like the C.T.A.\,\footnote{\textit{Cherenkov Telescope Array}.} array \cite{cta}.

The CT5 camera was delivered on site in Namibia in Spring 2012. Since then, the new telescope was fully integrated in the H.E.S.S. array and data are being accumulated to fully understand the camera response, the efficiency of the trigger decisions and the acquisition system performance. This testing phase consists of taking calibration runs (flat field runs, pedestal runs, SPE runs) as well as regular observation runs on selected ``benchmark'' targets. The results of this operating period will be described and discussed in the second part of this paper.

\section{Acknowledgements}

We would like to acknowledge the support of our host institutions. We also want to thank the whole H.E.S.S. collaboration for its support, and the referees, who helped us a lot improving the quality of the draft.

In addition, the LUPM team would like to thank Pierre-Eric Blanc, Sandrine Perruchot and Auguste Le Van Suu from the \textit{Observatoire de Haute-Provence}, who designed the calibration light source.

The support of the French Ministry for Research, the CNRS-IN2P3, the Astroparticle Interdisciplinary Programme of the CNRS and the \'Ecole Polytechnique is gratefully acknowledged.


\end{document}